\documentclass[useAMS,usenatbib,usegraphicx]{mn2e}
\usepackage{times}
\usepackage{graphicx}

\begin{document}
\newcommand{\goodgap}{%
 \hspace{\subfigtopskip}%
 \hspace{\subfigbottomskip}}

\title[Submillimetre emission from $\eta$ Carinae]
      {Submillimetre emission from $\eta$ Carinae}
\author[H.\ Gomez et al.]
{H. L.\ Gomez (n\'{e}e Morgan),$\! ^{1,2}$ L. Dunne,$^3$
  S. A.\ Eales,$\! ^1$ and M. G.\ Edmunds,$^1$ \\
$^1$ School of Physics \& Astronomy, Cardiff University,
     5 The Parade, Cardiff CF24 3YB, UK \\
$^2$Faulkes Telescope Project, Cardiff University, 5 The Parade, Cardiff CF24 3YB, UK \\
$^3$ School of Physics \& Astronomy, University of Nottingham, University Park, Nottingham NG7 2RD, UK}
\maketitle

\begin{abstract} 

We present critical, long-wavelength observations of $\eta$ Carinae in
the submillimetre using SCUBA on the JCMT at 850 and 450 $\mu$m to
confirm the presence of a large mass of warm dust around the central
star.  We fit a two-component
blackbody to the IR-submm spectral energy distribution (SED) and
estimate between $0.3-0.7\rm \, M_{\odot}$ of dust exists in the nebula
depending on the dust absorption properties and the extent of contamination from
free-free emission at the SCUBA wavelengths.  These results provide further evidence that
$\eta$ Carinae's circumstellar nebula contains $\rm > 10\,M_{\odot}$
of gas, although this may have been ejected on a longer timescale than
previously thought.

\end{abstract}

\begin{keywords}
circumstellar matter--stars: individual: $\eta$ Carinae--submillimetre
\end{keywords}

\section{Introduction} 
\label{intro} 

$\eta$ Carinae is one of the most luminous objects in our Galaxy ($\rm
10^6\, L_{\odot}$) and is well known for dramatic outbursts in which
material is ejected outwards from the star.  Its most famous mass
ejection is thought to have occurred in the early 19th Century, where
a few solar masses of gas was expelled in a few decades.  This feature
is clearly seen in observations as a bipolar nebula, known as the
Homunculus (Gaviola 1946).  The origin of the shape of the Homunculus
is unknown, but could occur due to ejection of mass in a non-uniform
interstellar medium i.e. if the ambient medium was denser in the
equatorial plane, the nebula would preferentially expand along the
poles.  Surrounding the Homunculus is a larger, less dense nebula,
known as the `outer ejecta' (e.g. Weis 2005) which contains numerous
filaments and dense condensates not dissimilar to those seen around
the young supernova remnant Cassiopeia A (e.g. Smith \& Morse 2004,
Fesen et al 2001).  The central star is hidden from view by this
structure but is thought to be a luminous blue variable (LBV) star
with mass $\rm 100\, M_{\odot}$ (Pittard 1999; Smith et al. 2003).
LBV stars are well known for large luminosities, violent instabilities
and periods of large mass loss.  The presence of a binary companion
(e.g. Pittard 2003) would provide an explanation for the catastrophic
mass ejections and the strong variabilities seen in the radio and
X-ray (Cox et al. 1995; Duncan \& White 2003).  The cycle of
periodicity suggests that any exisiting binary would have a highly
eccentric orbit, leading to strong wind-wind collisions such as those
seen in WR binaries (Marchenko et al. 2002; Pittard 1999).  The
evolutionary phase of an LBV star can last more than $10^4$\,yrs and
hence in this phase alone, a single star could lose roughly $\rm
1\,M_{\odot}$ of gas.  Episodic eruptions will increase the mass loss
further.  It is therefore not unreasonable to assume that the dust
mass loss rate could be $\rm \ge 10^{-6}\,M_{\odot}\,yr^{-1}$.  For
the lifetime of the average LBV phase, we could expect LBV stars to
inject $\ge 0.01\rm \,M_{\odot}$ of dust into the interstellar medium
and could be responsible for the dust emission seen around the young
supernova remnants Cas A (Dunne et al. 2003, Wilson \& Bartla 2005), Kepler (Morgan et
al. 2003), SN1998J (Pozzo et al. 2004), the Crab (Green, Tuffs \&
Pospescu 2004), SN2002 (Barlow et al. 2005) and more recently SN2003gd (Sugerman et al. 2006).

The structure of $\eta$ Car is revealed in the many multiwavelength
observations, yet there are issues surrounding the origin of the
Homunculus.  IR observations at 12 - 17\,$\mu$m revealed a small
structure within the Homunculus which appeared to be concentrated
around the `waist' of the nebula (Morris et al. 1999; hereafter M99).
They found a small bright core a few arcsecs across which they
interpreted as a disk/torus of dust with two components at 200\,K and
110\,K with $\rm \sim 0.15\, M_{\odot}$ of dust.  They suggested (with
a normal gas-to-dust ratio of 100) that the gas within this small
torus could therefore be $\rm \sim 15\, M_{\odot}$.  This argument is
flawed since estimating the 110\,K dust mass relied on observations at
the longer infrared wavelengths where resolution is limited to greater
than 1 arcminute, so the precise location of the 110K dust component
cannot be determined from the mid-IR images alone.

Davidson \& Smith (2000) also argued against the IR emission in a
small torus, proposing instead that this was too small to account for
the large luminosities radiated at 100\,$\mu$m.  The visual image
provides further evidence for this argument since there is no evidence
of an opaque dusty disk on such a small scale.  Hony et al. (2001)
then used high resolution images at 8 - 20\,$\mu$m to suggest that the
observed IR dust is not warm dust at 110\,K, but actually hot dust at
200\,K with a strange overlapping ring structure in the middle of the
nebula. They argue that these rings are similar to those seen in
SN1987A and are at different polar axes to each other.  If this
scenario is correct, the total dust mass in $\eta$ Car decreases to
$\sim \rm 10^{-4}\,M_{\odot}$ due to the strong dependence of dust
mass with temperature; the gas mass ejected in the Homonculus would
then be $< \rm 10^{-2}\,M_{\odot}$ and the evolution of $\eta$ Car
would not be as violent as currently thought.  This claim has been
disputed in Smith et al. (2000; 2003) who propose that the strange
smooth loop structure seen in the Hony et al. observations are simply
due to a disrupted torus.  They point to compact knots and hotspots in
the IR with filaments and arcs which could be due to a toroidal
distribution of dust which has been disrupted by subsequent mass loss.
They also use temperature maps from ISO data to show that cooler
`hidden' dust may exist in the outer polar lobes of the Homunculus and
not in a compact torus.

M99, Davidson \& Smith (2000) \& Smith et al. (2002) have all argued
that there is missing mass which has never been accounted for by the
observations, and the cooler dust mass which has been found has not
been resolved, so its precise location is unknown.  If there is
missing ejected mass, this will have important implications for the
energy budget and estimated parameters of LBV stars as well as its
future evolution. To determine if colder dust could exist further out
from the centre of the star, wavelength measurements are needed beyond
the mid-IR range (Davidson \& Smith 2000) which will dominate the dust
mass.  We therefore use archived SCUBA observations of $\eta$ Car to
investigate the distribution of the dust and the ejected mass in one of
the most studied objects in the nearby Universe.

\section{Submillimetre Observations} \label{sec:obs} 
\begin{figure}
\includegraphics[angle=-90,width=7.5cm]{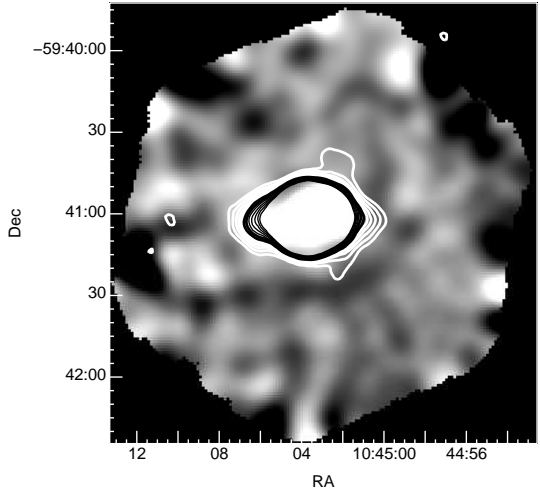}
\caption{\small{ $\eta$ Car at 450\,$\mu$m with 850\,$\mu$m
contours overlaid. White contours start at 5-$\sigma$ (160 mJy/beam) incrementing by 2-$\sigma$ with black contours starting at
10-$\sigma$.  Both
images have been smoothed to the same resolution and are in J2000
coordinates.  }} 
\label{eta} 
\end{figure}

The sub-mm
observations of $\eta$ Carinae were originally observed by SCUBA on
the JCMT in January 1998 in Grade II weather, simulataneously
observing at 450 and 850 $\mu$m (with nominal beam size of 8 and 14
arcsec respectively).  The data was taken from the JCMT archive
(originally observed by Henry Matthews, program ID M97bc30) and consisted of a single
Jiggle map observation with a 120 arcsec chop throw in Alt-Az and
position angle $90^{\circ}$. This was reduced using the standard data
reduction pipeline SURF (Sandell et al. 2001) with sky opacities
obtained from CSO-FITS and skydips to give relatively stable optical
depths of $\tau_{450}\sim 0.8$ and $\tau_{850} \sim 0.2$ (see Section~\ref{sec:ext}).  The
remsky option was used with the outer ring of bolometers flagged as
sky bolometers (as $\eta$ Car is almost a central point source in the
maps) and the option for addback was turned on.  

\begin{figure*}
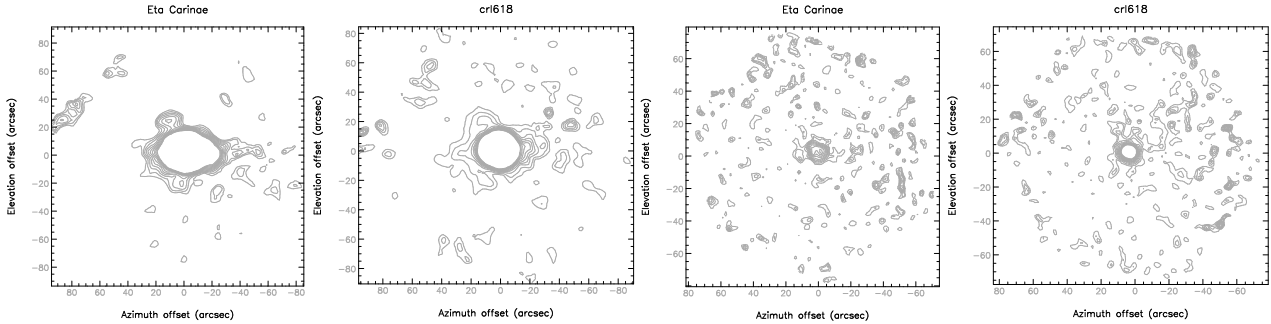

\includegraphics[angle=-90,width=4cm]{fig2.ps}
\includegraphics[angle=-90,width=4.3cm]{fig3.ps}
\includegraphics[angle=-90,width=4cm]{fig4.ps}
\includegraphics[angle=-90,width=4.3cm]{fig5.ps}
\caption{\small{ (a) Contour plot of $\eta$ Carinae at 850\,$\mu$m displaying 1 percent of the peak voltage up to 10 percent.   (b) Contour levels for the calibrator CRL618 showing the same levels.  Extension along the midplane is seen in both calibrator and source. (c) $\eta$ Carinae at 450\,$\mu$m.  Contours starting at 15 percent of the peak voltage up to 95 percent.  (d) CRL618 at 450\,$\mu$m.  Contour levels 5 percent of the peak voltage up to 50 percent.
}} 
\label{etasubmm} 
\end{figure*}

The 450\,$\mu$m SCUBA image of $\eta$ Car is shown in Fig.~\ref{eta} with 850\,$\mu$m
contours overlaid. White contours start at 5-$\sigma$ (160 mJy/beam) incrementing by 2-$\sigma$ with black contours starting at
10-$\sigma$.  The central part is
not resolved by the SCUBA field-of-view but we see an interesting
feature which appears to be extended in the x direction, along the
midplane of the image which is particularly clear at 850\,$\mu$m.  It is difficult to determine the significance of this since the source
is at such a high airmass. It is well known that
larger chop throws can cause `smearing' along the direction of the
throw, in this case the x-axis, and the image of $\eta$ Car is similar
to example images of this problem\footnote{see
http://www.jach.hawaii.edu/JCMT/continuum/}. 
Extended emission 
along the midplane is also seen in the calibrator CRL618 at 850\,$\mu$m (Figs~\ref{etasubmm} (b) \& (d)), suggesting it may be a combination of the chopping method and the beam pattern at high airmass.   The extension is not as clear in the 450\,$mu$m image of the calibrator.   Radial profiles of both the beam (taken from the calibrator) and the source show that the extension cannot be separated from the beam pattern.

\section{Calibration \& Errors}

The high airmass of $\eta$ Carinae ($\sim 5.5$) at the
time of observation produces potential difficulties with 
calibration which we will address carefully in this section. The calibrator observed on the same night was at a lower airmass of 2.5. Calibration of the data requires two steps, firstly an extinction
correction for the absorption of incoming radiation by the atmosphere,
and secondly a comparison with a calibrator of known flux to convert
instrumental units (V) into Jy. 

\subsection{Extinction correction:} 
\label{sec:ext}
For the time the data was taken, the JCMT has two methods available to
measure the value of the zenith sky opacity ($\tau$) at the
wavelengths of interest; skydips and extrapolation from the CSO 1.1 mm
$\tau$. Skydips are taken at the SCUBA wavelengths (450 and 850\,$\mu$m)
but are only made infrequently (every two hours or so). The CSO
radiometer measures the $\tau$ at 1.1 mm every 10 minutes, and so
samples changes in the atmosphere on a much more useful timescale. The
JCMT have created relationships between the CSO $\tau$ values and
those at 450 and 850\,$\mu$m using an extensive data-base of
skydips. They also fit the CSO $\tau$ data for each night with a
polynomial which allows interpolation between measurements and reduces
the effect of any `spikes' in the CSO measurements. These polynomial
fits are the recommended way to extinction-correct SCUBA maps. 

\begin{figure}
\centerline{\includegraphics[width=8cm,height=7cm]{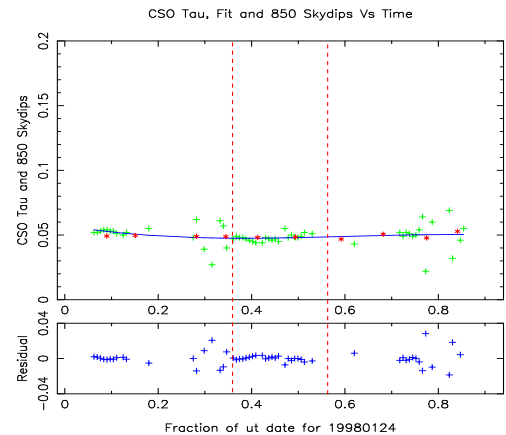}}
\caption{{\label{csotauF}}\small{
The CSO tau, polynomial fit and 850\,$\mu$m skydips for UT
24/01/1998. The residuals are shown below. The times used to derive
the uncertainty in $\tau$ are shown by dashed lines, $\eta$ Car was
observed at $\sim 0.5$ on the horizontal axis. The larger scatter in
$\tau$ values at UT fractions of 0.3 and 0.8 correspond to sunset and
sunrise, where the sky becomes more unstable. As both our calibrator
and object were observed shortly after midnight, we have only taken
the stable portion of the night to estimate $\sigma_\tau$.}}
\end{figure}

On the night the data were taken, the sky was extremely stable. The plot of the CSO $\tau$ values, the polynomial fit and the residuals is shown in Fig.~\ref{csotauF}. The part of the night considered to be `stable' after sunset and before sunrise (when the $\eta$ Car data were taken) is shown between the dashed lines.

 The value of $\tau$ which is used to correct the flux of the object
 for atmospheric absorption, is uncertain - and this uncertainty,
 $\sigma_\tau$, leads to an uncertainty on the flux measured. Due to
 the high airmass of $\eta$ Car, any error in the $\tau$ value is
 amplified greatly when applied. From Dunne \& Eales (2001), we
 have the fractional error in flux due to the uncertainty in $\tau$
 as:

\begin{equation}
\frac{\Delta F}{F} = 1-e^{-A\sigma_\tau}  \label{fracfluxE}
\end{equation}

where we have taken $\sigma_\tau$ to be the r.m.s. of the residuals in
Fig.~\ref{csotauF}, and $A$ is the airmass.  The value of
$\sigma_\tau$ is 0.00253 and this must be translated to an r.m.s. on
the $\tau$ values at 450 and 850\,$\mu$m. For this we use the
pre-upgrade revised narrow-band filter relationships between CSO tau
and 450/850\,$\mu$m given on the JCMT
webpages\footnote{http://www.jach.hawaii.edu/JCMT/continuum/calibration/atmos/tau.html},
and using propagation of errors we find that $\sigma_\tau$(850) = 0.005  and $\sigma_\tau$(450) = 0.012.

Using Eqn.~\ref{fracfluxE} we calculate the fractional error on the
flux due to uncertainties in the
extinction corrections as $\sigma_{ext}$ (850) = 3 per cent and $\sigma_{ext}$ (450) = 6 per cent.
If we instead take the extreme values of the residuals rather than the
r.m.s. we get increased values of 6 (850) and 15 per cent (450), however,
these are not statistically correct, and we include them as merely
illustrative of a worst case.

\subsection{Changes in the gain:}

The second part of the calibration procedure is to compare the signal
measured in an aperture on the object with that in the same aperture
on a calibrator of known flux. The calibrator used was CRL618, chosen
because it was observed shortly before $\eta$ Carinae, and therefore sky
conditions and the shape of the dish would be closest to those at the
time of the $\eta$ Car observation. A change in the dish shape (due to
thermal relaxation after sunset), or a telescope which is poorly
focussed will result in changes in beam shape and hence the gain -
defined as the ratio of calibrator flux (Jy) / calibrator flux
(V). Thus, if the calibrator is observed under as similar conditions
as possible to the objects then this effect is minimised.

A check of all of the calibration maps during the night showed that
the beam shape was very poor in the earlier part of the evening (after
sunset), despite many attempts at focussing. However, for the
calibrator map taken prior to $\eta$ Car, alignments were also made on
the secondary mirror in the x-y plane which resulted in a much
improved beam shape. Thus we believe that the previous poor quality
images were due to a poorly aligned mirror and there is no reason to
believe that this will affect the source data which was taken just
after the re-alignment.

There could be some concern that observing at such low elevations may
also affect the shape of the dish and hence the gains, however, the
staff at the JCMT have found no evidence for a link between gain and
elevation (Iain Coulson, private communication). Furthermore, the beam
shapes look similar between the calibrator and $\eta$ Car (Fig.~\ref{etasubmm}), again
suggesting that there is not a serious change in images observed at the lower
elevations. Any deterioration in dish shape tends to lead to higher
gains (more Jy per V) and therefore any elevation effect on the dish
would be leading us to underestimate the flux of $\eta$ Car rather
than overestimating it (as the calibrator was observed at a lower
airmass).

Since the earlier calibration maps had been affected by the
mis-alignment of the secondary mirror, and therefore do not reflect
usual nightly variations in the gain, we will use the information from
Tables 1 \& 2 in Dunne \& Eales (2001) to estimate the uncertainty on
the gain. Here the variation of gain values in various sized apertures
was calculated using data taken over the same period as these
observations (July 1997 - July 1998). This gives the uncertainty in
the gain in a 45 arcsec aperture as $\sigma_{gain} = 6$ per cent at 85\,0$\mu$m and $\sigma_{gain} = 9$ per cent at 450\,$\mu$m.  

There will also be an absolute calibration uncertainty due to errors
in the flux of the calibrator and the model used to generate the
planet fluxes (on which the flux of the secondary calibrators is
based).  The errors on the flux of CRL618 are quoted as $4.7\pm 0.12$ Jy at
850\,$\mu$m and $11.8\pm 0.6$ Jy at 450\,$\mu$m. Adding in quadrature an
additional 5\% error due to the uncertainty in the model, the total
error on the calibrator flux is estimated as $\sigma_{abs} = 5.6$ per cent at 850\,$\mu$m and
$\sigma_{abs} = 7.0$ per cent at 450\,$\mu$m.  

The total calibration error is then the quadratic sum of the three
terms for extinction, gain and absolute as follows:

\begin{equation} 
\sigma_{cal} = \sqrt{\sigma_{ext}^2 + \sigma_{gain}^2 + \sigma_{abs}^2}
\end{equation}

\noindent giving $\sigma_{cal}$(850) = 9 per cent and $\sigma_{cal}$(450) = 13 per cent.
While these numbers may look surprisingly low, it must be remembered
that this was a night with exceptionally good observing conditions
(during the El Nino period), and that the best practise has been used
in the calibration procedure.

\subsection{The SCUBA Fluxes}

The integrated flux was measured by placing the same aperture over
both $\eta$ Car and CRL618, measuring the total flux in
Volts on both.  The fluxes of $\eta$ Car at 850 and 450\,$\mu$m were
found to be $12.8 \pm 0.8$\,Jy and $52 \pm 6.8$\,Jy respectively as measured in an aperture of
diameter $45^{\prime \prime}$. 

The radius of the aperture needed to incorporate the entire emission
is $\sim 28^{\prime \prime}$, which at the distance of $\eta$ Car
corresponds to a physical size of 0.3\,pc and is just larger than the
visual size of the Homunculus.  The dust seen in these images may be
probing the `outer ejecta' region (e.g. Smith \& Morse 2004, Weis
2004).  However, we cannot resolve the inner $10^{\prime \prime}$ of
the Homunculus and we refer the reader to the recent sub-arcsecond
study of $\eta$ Car e.g. Chesneau et al (2005).

\section{The IR-Radio Spectral Energy Distribution} 
\label{sec:dust} 

Due to the high variability in the X-ray, Radio, mm and IR and as the
sub-mm observations of $\eta$ Car were taken during a shell-event in
1998, we chose literature fluxes taken nearest this epoch
(corresponding to a radio quiet stage).  This allows for the best possible direct
comparison of the sub-mm with any variability in the radio, as well as determining the evolution of the SED during
this time (e.g. Abraham \& Damineli 1999; Abraham et al. 2005).
Where unavailable, older literature fluxes were used (Cox
et al. 1995).  The 1992 millimetre fluxes taken from this work are thought to be at a similar level to those expected in 1998 (e.g. Fig 1, White et al. 2004).

\begin{figure*}
\includegraphics[angle=-90,width=11cm]{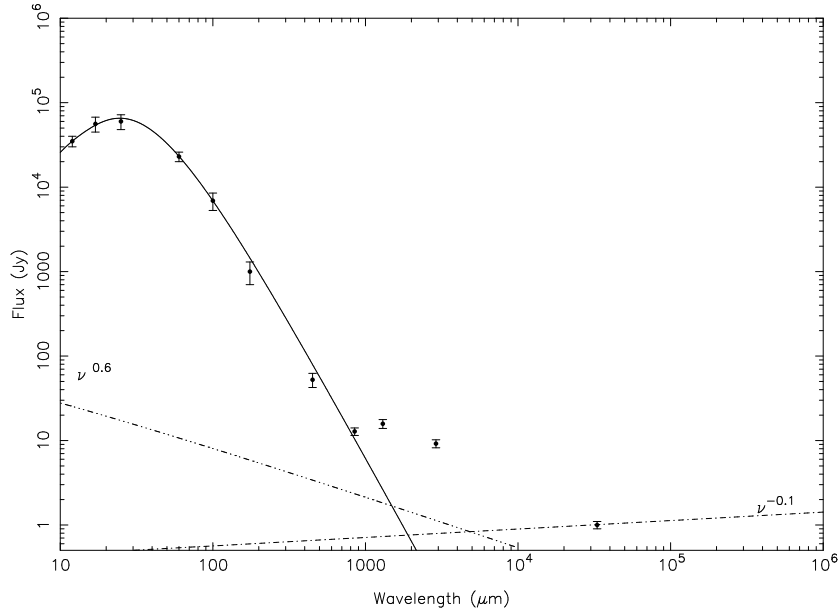}\hfill
\caption{\small{The spectral energy distribution from 12 -
850\,$\mu$m through to the radio regime of $\eta$ Carinae for the 1998 epoch.  The 175\,$\mu$m data point is also included although this was observed in 1977 (Harvey, Hoffman \& Campbell 1978). The solid line is the two temperature
component fit to the infra-red fluxes.  The dashed line shows the SED expected
from theoretical free-free emission (Lamers \& Cassinelli 1999) with wind and stellar parameters taken from
Humphreys (1988) and Pittard (1999).  This falls short of the observed fluxes at the mm wavelengths from Cox et al. (1995).  The dot-dashed line represents
the emission from the optically thin, ionised Homunculus with spectral
slope $\alpha=-0.1$ normalised to the 3\,cm flux (Cox et al. 1995).  
}}
\label{etalow450} \end{figure*}

We fitted the SED of $\eta$ Car from 10 - 850\,$\mu$m using a modified
blackbody (Fig.\ref{etalow450}) requiring a two temperature component
at around 200 and 110\,K. The dust emissivity exponent $\beta$, is $< 1.5$,
lower than the nominal value of 2 for `normal' interstellar
composition but higher than those estimated in M99 and Smith et al
(2003).  This indicates that the grains could be amorphous, large or
non spherical in composition (e.g. Bohren \& Huffmann 1983). However,
there is likely to be contamination from the radio emission at the
sub-mm wavelengths which is dependant on the phase of the radio cycle.
There are two sources of radio emission from the star - free free
ionised emission from the stellar wind and emission from the optically
thin Homunculus region (Cox et al. 1995).  Both thermal processes
depend on frequency, free-free varies as $\nu^{0.6}$ (Wright \& Barlow 1975) and the optically thin region as
$\nu^{-0.1}$. The contamination from the radio to the sub-mm emission
can be removed by determining the flux expected at 450 and
850\,$\mu$m.  The variation of free-free emission in Jy with
wavelength from an ionised stellar wind is given in Lamers \&
Cassinelli (1999). X-ray modeling of colliding wind binaries fit the
observations of $\eta$ Car with parameters: mass $\sim \rm
70~M_{\odot}$, $\rm T_{eff}\sim 20,000K$, mass loss rate, $\rm \dot{M}
\sim 9 \times 10^{-5}~M_{\odot}~yr^{-1}$, terminal velocity $\rm \sim 9300~ km~s^{-1}$ and distance 2.6 kpc (Pittard 1999).
The SED of the free-free emission assuming these values and a doubly
ionised wind is shown in Fig.~\ref{etalow450} (dashed line). The power
law is {\it significantly} below the sub-mm fluxes and also underpredicts
the observed 1.2 and 3\,mm fluxes.  Cox et al. (1995) suggest that the
power law slope from the free-free emission is actually $\nu^{1.0}$ in
the mm regime due to hydrogen recombination changing the electron and
proton densities in the wind (see their Fig 3).  The emission then turns over at
$\lambda \sim 1.2$\,mm to the normal slope of $0.6$, just below the observed flux at this wavelength, shown on Fig~\ref{etalow450}. 
We use the {\it observed} fluxes rather than the theoretical prescription of a stellar wind to
estimate the maximum possible contribution from the ionised stellar
wind in the sub-mm regime.   Thus we would expect a flux of $\sim 12$ Jy at 850 and 18 Jy at 450\,$\mu$m respectively due to the ionised stellar wind.  It is clear from Fig~\ref{etalow450} that almost all of the 850\,$\mu$m flux could be free-free emission, yet not all of the 450\,$\mu$m emission can be explained from the expected power law.  Indeed, it is difficult to explain the 450 and 850\,$\mu$m flux ratio with a radio emission process.   However, given the uncertainty of how much of the sub-mm flux is due to dust emission or contaminated by free-free, we have two scenarios for the dust mass in $\eta$ Car (1) best case: all of the estimated SCUBA flux is due to dust and (2) worst case: 94 and 35 per cent of the 850 and 450\,$\mu$m flux respectively is due to the ionised stellar wind.  This effectively gives a lower limit to the range of dust masses expected (see Section~\ref{sec:outer}).  

As noted in Smith et al. (2003), the fit to the energy distribution is
not unique, a range of values of temperature could achieve a
satisfactory fit to the spectrum within the errors on the fluxes.
Indeed in their work, they reproduce an adequate fit to the Morris et
al. (1999) data with dust temperatures of 140\,K and 200\,K.  However,
none of the previous best-fit SEDs produce the correct 450/850 ratio
as observed here.  The extra fluxes at the long wavelength tail
provide another constraint, which forces the fit to have a colder
temperature than
  the previous models at 110\,K.  It is extremely
  important to separate the
  dust emission from free-free at these wavlengths as this affects the
  dust mass in the nebula.  If the wind is optically thin, for
  example, most of the IR emission would be due to dust and not
  contaminated by free-free.

\section{A Large Mass of Dust?}
\label{sec:outer}

\begin{figure}
\includegraphics[angle=-90,width=9cm]{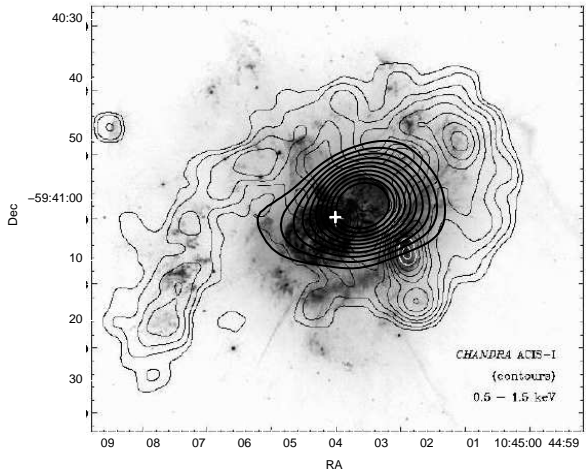}\hfill
\caption{HST/WFPC2 image taken from Smith \& Morse (2004) with
contours of soft X-ray emission (thin black lines) and smoothed 450\,$\mu$m
(thick black lines).  The submm contours start at 3-$\sigma$ and
extend just outside the Homunuculus region shown by the well known HST
bi-polar nebula and may be associated with the outer ejecta. Indeed there is a good correlation between the location of the 450\,$\mu$m emission and the hard X-Rays from the shocked wind.  The
image was aligned using the position of the central star.  Co-ordinates are J2000.}
\label{etahst} \end{figure}
The emission seen in the SCUBA
 images and the radial profiles suggests that some of the dust may be located outside the
 Homunculus although due to the beam pattern we cannot determine if this is  a real extension.  The HST/WFPC2 image is shown in Fig.~\ref{etahst} with
 the {\small CHANDRA} X-ray observations and the submm emission shown
 as contours starting at S/N $>$ 3-$\sigma$ (original image from
 Smith \& Morse 2004).  The 450\,$\mu$m emission
 appears to coincide with the X-ray emission which is thought to occur
 where previous mass loss from the star is colliding with the
 surrounding interstellar material, causing shock heating of the
 gas. We may therefore be probing dust shells from a previous
 mass-loss phase\footnote{A rough calculation shows that the dust
 could not be due to swept up from interstellar gas since the swept up dust
 mass at this distance would be
 less than $\rm < 10^{-3}\,M_{\odot}$.}.  

To determine the dust mass in $\eta$ Car, we use \begin{equation}
M_d={S_{\nu}D^2 \over{\kappa_{\nu} B_{\nu}(\nu , T)}}
\label{dust_mass} \end{equation} where $\kappa_{\nu}$ is the dust mass
absorption coefficient ($\sim 0.27\rm \,kg^{-1}\,m^2$ for normal
interstellar dust at $\lambda_{ \rm 450\mu m}$) and D is the distance
to the star.  The dust mass from the best fit SED to the free-free subtracted fluxes (scenario 2) is $\rm <
0.54\pm 0.1\,M_{\odot}$ depending on the absorption properties of the dust.
 With an typical gas-to-dust ratio of 100, this suggests that
 around $\rm 50\,M_{\odot}$ would have been expelled in the wind.
 Such a large number may indicate that the gas-to-dust
 ratio in the atmospheres of massive stars could be higher than
 predicted (e.g. AG Car - Voors et al. 2000). Alternatively, the dust
 could be more efficient at emitting than normal ISM dust.  Previous
 modelling of the SED of $\eta$ Car has shown that the dust does not
 follow an opacity law of $\kappa \propto \lambda^{-\beta}$ (Robinson
 et al 1987).  If we use the absorption coefficient estimated
 from Dunne et al. (2003) to represent pristine, newly formed dust in
 stellar winds, we obtain $\rm 0.3\,M_{\odot}$ which would require
 $\rm \sim 30\,M_{\odot}$ of gas to be expelled.  Using the
 best fit fluxes {\it without} the free-free contamination (scenario 1), the dust
 mass would be $\rm 0.7 \pm 0.1\,M_{\odot}$.   Errors quoted are given using 68 per cent confidence intervals from parameters obtained using the bootstrap technique (whereby 1000 sets of artificial fluxes were created from the original data with each individual set then fitted with our two-temperature model).

This is in good
 agreement with the original hypothesis of M99 and
 Smith et al. (2003).  The dust mass calculated here 
 appears to be spread out over a larger area than the torus and
 quite possibly outside the polar lobes of the Homunculus.  Combining
 this result with that in M99 and Smith et al.  (2003), we obtain the
 dust masses for
 each component as listed in Table~\ref{etacomp}.  Since we cannot
 resolve the inner Homunculus, the dust mass here in the outer ejecta
 is likely to include the mass postulated to be in the lobes or
 the torus. 

\begin{table} \begin{tabular}{ccccc} \hline \hline
\multicolumn{1}{c}{Region}&\multicolumn{1}{c}{Size ($^{\prime
\prime}$)} &\multicolumn{1}{c}{Temperature}&\multicolumn{1}{c}{Dust
Mass ($\rm{M_{\odot}}$)}\\ \hline 
Core & 0.3 & 400 & $3 \times
10^{-4}$$^{*}$\\ 
Lobes, torus & 5 & $200^{*}$ & $0.02^{*}$\\ 
Lobes & 20 &$140^{*}$ & $0.13^{*}$ \\ 
Lobes \& Outer ejecta? & 28 & 114 & $0.54~\pm 0.1^{**}$ \\ 
\hline \end{tabular}
\caption{\label{etacomp}\small{Components fitted to the IR-submm
SED at various temperatures.  $^*$Average results published in
   M99 and Smith et al. (2003) are included for
   completion.  They
   used a dust emissivity of 1 - 1.2 in their work. $^{**}$ Dust mass obtained using the best-fit parameters to the SED with the free-free subtracted SCUBA fluxes, scenario (2). Errors are 68 per cent confidence intervals estimated using the bootstrap technique.}} \end{table}

If we assume that the dust around $\eta$ Car is similar to dust in the
ISM, then the Homunculus and the surrounding area contains roughly
$\rm 0.5,M_{\odot}$ which has been ejected during the
 the last $10^4$\,yrs. The averaged dust mass loss rate over the
 entire phase is then $\rm \sim 5 \times 10^{-5}\,M_{\odot}\,yr^{-1}$.  This is
 an order of magnitude greater than the dust mass loss rate estimated
 from observations of Wolf-Rayet binaries (Marchencko et al. 2002) and
 more than suggested by previous IR observations of $\eta$ Car.  If
 the submm dust is located in the polar lobes of the nebula
 then it must
 have been ejected during the same singular event which caused the
 lobes - the great eruption.  This eruption lasted only a few decades
 which gives a large time-averaged gas mass loss rate of $\rm >
 0.5\,M_{\odot}\,yr^{-1}$ (for more efficient dust emitters).  Such a
 high mass loss rate was first noted in Smith et al. (2003) and
 suggests that the pulsational mass loss in $\eta$ Car is far more
 extraordinary than imagined. 

 The sub-mm observations suggest that an upper limit of $\rm
0.7\,M_{\odot}$ of dust could exist in the extended dust shells, dust
which must have been formed in the stellar wind with effective
temperatures $\rm T_{eff} \gg 2000\,K$.  This indicates that contrary
to the study in Morgan \& Edmunds (2003), significant amounts of dust
could form in the stellar winds of massive stars even when the
conditions in the atmospheric envelope are not reached by the
available hydrodynamical models.  This previously undetected mass of
material in the outer ejecta has
 important consequences for our understanding of stellar evolution and
 the formation of dust in massive stars.
 Massive stars may therefore provide a sustainable dust yield to the
 ISM, in which case their contribution in chemical evolution models
 could be severly underestimated (e.g. Whittet 2003).

We have used critical long wavelength observations to show that a
large mass of dust does exist around $\eta$ Carinae and suggest that
the dust is distributed over a larger region than previously thought.
This has important implications for stellar evolution and may provide
a natural explanation for the origin of the dust seen in young
supernova remnants.  However $\eta$ Car is extremely
massive and highly unstable, so the huge dust mass loss rate is not
unexpected.  It is very unlikely that the lower mass stars which were
responsible for the supernova events of Kepler and Cas A would have
ejected enough mass to form such a large quantity of dust before the
explosion. In order to determine a more accurate dust mass in
$\eta$ Car, we require simultaneous observations in the sub-mm and mm
to separate the contamination of the free-free emission.  We also require better resolution to determine how extended the sub-mm emission is. Future
observations with Herschel and SMA will allow us to overcome this problem.   These observations should also allow us to determine any
variability of the star in the submm.  Quantifying the dust injection from
stars of different masses in different stages of their evolution will
have important consequences for estimating the dust budget in both
local Galaxies and those at high redshifts as well as providing
fundamental information about stellar evolution.

\section*{Acknowledgements}

HLG is a research fellow of the Royal Commission for the Exhibition of
1851. We thank Iain Coulson and Jan Wouterloot for help with the data reduction and
Nathan Smith for interesting and informative discussions.  We also thank the referee for constructive comments which have improved the paper.  The James Clerk Maxwell Telescope is operated by The Joint Astronomy Centre on behalf of the Particle Physics and Astronomy Research Council of the United Kingdom, the Netherlands Organisation for Scientific Research, and the National Research Council of Canada.


\begin{thebibliography}{} 
\bibitem{} Abraham Z., Damineli A., in {\it
Eta Carinae At The Millennium}, ASP Conference Series vol. 179, 1999
eds. J. A. Morse, R. M. Humphreys, and A. Damineli, p 263
\bibitem{}Abraham Z.,Falceta-Gon\c{c}alves D., Dominici T. P., 
Nyman L.-\AA.,
Durouchoux P., McAuliffe F., Caproni A., Jatenco-Pereira V.,
 2005, A \& A, 437, 977
\bibitem{}Barlow M.J., et al. 2005, ApJL, 627, 113
\bibitem{}Bohren C.F., \& Huffman D.R., 1983, Absorption and Scattering of 
Light by Small Particles, Wiley, New York
\bibitem{} Chesneau O., Min M., Herbst T., Waters L.B.F.M., Hillier D.J.,
Leinert Ch., de Koter A., Pascucci, I., et al., 2005, A \& A, 435, 1043 
\bibitem{}Cox P., Mezger P.G., Sievers A., Najarro F., Bronfman
L., Kreysa E., Haslam, G., 1995, A \& A, 297, 168 
\bibitem{} Davidson K. \& Smith N., 2000, Nature, Vol 405, P.532
\bibitem{} Duncan R.A., \& White S.M., 2003, MNRAS, 338, 425
\bibitem{} Dunne L., Eales S., 2001, MNRAS, 327, 697 
\bibitem{} Dunne L., Morgan H.L., Eales S.,
Ivison, R., Edmunds, M.G., 2003, Nature, 424, 285
\bibitem{} Dunne L., Eales S., Edmunds M.G., Ivison R., Alexander,
P., Clements D.L., 2000, MNRAS, 315, 115
\bibitem{} Fesen R.A., Morse J.A., Chevalier R.A., Borkowski K.J.,
Gerardy C.L., Lawrence S.S., van den Bergh S., 2001, AJ, 122, 2644
\bibitem{} Gaviola E., 1946, Rev. Astron., 18, 252
\bibitem{} Green D., Tuffs R.m., Popescu C., 2004, MNRAS, 355, 1315
\bibitem{} Harvey P.M., Hoffman W.F., Campbell M., F., 1978, A \& A, 70, 165
\bibitem{}Hony S., Dominik C., Waters L.B.F.M., Icke V.,
Mellema G., van Boekal, R., de Koter A., Morris P.M., Barlow M.,
Cox P., Kaufl H.U., 2001, A \& A, 377, L1
\bibitem{}Lamers H.J.G.L.M., \& Cassinelli J.P., 1999, {\it An
Introduction to Stellar Winds}, 46, Cambridge University Press, Great
Britian
\bibitem{}Marchenko S.V., Moffat A.F.J., Vacca W.D., C\^{o}t\'{e}, S.,
Doyon R., 2002, ApJ, {\  565}, L59
\bibitem {}Morgan H.L., Edmunds M.G., 2003, MNRAS, 343, 427
\bibitem{} Morgan H.L., Dunne L., Eales S., Ivison R., Edmunds,
M.G., 2003, ApJ, 597, L33
\bibitem{} Morris. P.W., et al., 1999, Nature, 402, 502 (M99)
\bibitem{} Pittard J., PhD Thesis, 1999, University of Birmingham
\bibitem{} Pittard J., 2003, Astronomy and Geophysics, Vol 44, P.1.17-1.22
\bibitem{} Pittard J.M., Stevens I.R., Corcoran M.F., Ishibashi K., 1998, 
MNRAS, 299, L5
\bibitem{} Polomski E., Telesco C.M., Pina R.K., Scott Fisher, R.,
1999, ApJ, 118, 2369
\bibitem{} Pozzo M., Meikle W.P.S., Fassia A., Geballe T., Lundqvist
P., Chugai N.N., Sollerman J., 2004, MNRAS, 352, 457
\bibitem{} Robinson G., Mitchell R.M., Aitken D.K., Briggs G.P., Roche P.F., 1987, MNRAS, 227, 535
\bibitem  {} Sandell G., Jessop N., Jenness T., 2001, SCUBA Map Reduction
Cookbook, starlink cookbook 11.2 
\bibitem{} Smith N., Gehrz R., 2000, ApJ,529, L99 
\bibitem{} Smith N., Gehrz R.D., Hinz P.M., Hoffman W.F., Mamajek,
E.E., Meyer M.R., Hora J.L., 2002, ApJ, 567, L77
\bibitem{} Smith N., Gehrz R.D., Hinz P.M., Hoffmann W.F., Hora
J.L., Mamajek E.E., Meyer M.R., 2003, ApJ, 125, 1458
\bibitem{}Smith N.M., Morse J.A., 2004, ApJ, 605, 854
\bibitem{} Sugerman B.E.K., Ercolano B., Barlow M.J., Tielens A.G.G.M., Clayton G.C., Zijlstra A.A., Meixner M., Speck A., et al., 2006, Science, 313, 196, pre-print, astro-ph/0606132
\bibitem{} Voors R.H.M., et al., 2000, A \& A., 356, 501
\bibitem{} Weis K., {\it in The Fate of the Most Massive Stars}, 2005
ASP Conference Series, Vol. 332, eds. R. Humphreys and K. Stanek, p.275
\bibitem {} White S.M., Duncan R.A., Lim J., Nelson G.J., Drake S.A., Kundu M.R., 1994, ApJ, 429, 
380
\bibitem{} White S.M., Duncan R.A., Chapman J.M., Koribalski B., 2005, {\it in The Fate of the Most Massive Stars},
ASP Conference Series, Vol. 332, eds. R. Humphreys and K. Stanek, p.129
\bibitem{} Whittet D.C.B., 2003, Dust in the Galactic Environment
Second Edition, IOP, Cambridge University Press, UK
\bibitem{} Wilson T.L., Batrla W., 2005, A\&A, 430, 561
\bibitem{} Wright A.E., \& Barlow M.J., 1975, MNRAS, 170, 41  

\end{thebibliography}
\end{document}